\documentclass[conference]{IEEEtran}

\ifCLASSINFOpdf
  
\else
 
\fi

\usepackage{graphicx}
\begin{document}

\title{Memristor-based Approximation of Gaussian Filter}

\author{
\IEEEauthorblockN{Alex Pappachen James, Aidyn Zhambyl, Anju Nandakumar}
\IEEEauthorblockA{Electrical and Computer Engineering department\\Nazarbayev University School of Engineering\\
Astana, Kazakhstan\\
Email: alex.james@nu.edu.kz, aidyn.zhambyl@nu.edu.kz, anju.nandakumar@nu.edu.kz}
}

\maketitle

\begin{abstract}

Gaussian filter – is a filter with impulse response of Gaussian function. These filters are useful in image processing of 2D signals, as it removes unnecessary noise. Also, they could be helpful for data transmission (e.g. GMSK modulation). In practice, the Gaussian filters could be approximately designed by several methods. One of these methods are to construct Gaussian-like filter with the help of memristors and RLC circuits. Therefore, the objective of this project is to find and design appropriate model of Gaussian-like filter, by using mentioned devices. Finally, one possible model of Gaussian-like filter based on memristor designed and analysed in this paper.  
\end{abstract}\


\section{Introduction}
Gaussian filter – in signal processing a filter with impulse response of Gaussian function. This type of filters has no overshoots to input of a step function while minimizing fall and rise time. Mathematically this filter applies convolution of input signal with Gaussian function. Gaussian filters used in image processing for two dimensional images and Gauss modulation {\cite{1}}. These filters widely used to limit the spectral energy which are out of transmission band in digital communications. Therefore, this type of filter is usually implemented in digital domain. However, they are also used to model analog delays. Thus, Gaussian filters should be implemented in analog domain, which requires faithful approximation of the filter {\cite{2}}.\\
		
	On the other hand, memristor (memory resistor) – electronic device, which is known as the fourth fundamental element after resistor, capacitor and inductor. Its resistance is known as memristance, which is the function of current passing through it and flux. Scientist argue that the use of memristor seems more beneficial, as it consumes less energy, produces less heat, do not consume power when idle and comfortable with CMOS interfaces {\cite{3-4}}. Today, the implementation of memristors to electrical circuits of chips seem highly sensitive.\\
		
	The aim of this paper is to design a filter, which operate as Gaussian filter. The filter must be designed via using RLC devices and with the implementation of memristors to the designed circuit. In the second section of this paper backgrounds about Gaussian filter and memristors is going to be provided. The objective of the third section is to find out appropriate model by simulating different circuits. Then in the fourth section some mathematical analysis will be provided to obtained results and finally the paper will be concluded by stating the result of the done work.

\section{Gaussian Filter and Memristor}
\subsection{Gaussian Filter}

	Gaussian filter is such filter which convolves the input signal with the impulse response of Gaussian function. In some sources this process is also known as the Weierstrass transform {\cite{2}}. The Gaussian function is given as in equation 1, where $\mu$ is the time shift and $\sigma$ is the scale. For input signal of $x(t)$ the output is the convolution of $x(t)$ with Gaussian, as shown in equation 2.
\begin{equation}
f(x)=\frac{1}{\sigma \sqrt{2\pi}} \exp{-\frac{(x-\mu)^2}{2 \sigma^2}}
\end{equation}\\
\begin{equation}
\int_{-\infty}^{\infty}{x(t) f(\tau-t)d\tau}
\end{equation}\\
As a signal $x(t)$ passes through a system $H(s)$, the impulse response is also $H(s)$. Hence, if this impulse response is a Gaussian function, then transfer function $H(s)$ approximates required Gaussian. Authors in the mentioned paper ({\cite{2}}) mathematically solve for $H(s)$ via using the technique known as $Levin's$ \ $t-transform$. This method finally gives equation 3 for transfer function of a fourth order Gaussian filter. Therefore, by inspecting, proper values for $\sigma$ and $\mu$ could be selected such that $H(s)$ approaches Gaussian. 
\begin{equation}
H(s)=\frac{a_3 s^3 + a_2 s^2 + a_1 s + a_0}{b_4 s^4 + b_3 s^3 + b_2 s^2 + b_1 s + b_0}
\end{equation}\
\subsection{Memristor}
Memristor (memory resistor) is an electrical device such as resistor, capacitor and inductor. The units of resistance of memristor is defined in ohms. It is a passive circuit element which mathematically could be defined as a relationship of the time integrals of current and voltage across two terminals of this device. Scientists in {\cite{5}} states that the resistance of memristor change as the direction of current passing through it changes; namely: the resistance increases as the current flows in one direction and decreases as the current flows in opposite direction {\cite{10}}. Furthermore, it is stated that memristor “remembers” the current that flowed before through it. To define memristance, equations 4-6 could be written.

\begin{equation}
d \Phi = M dq
\end{equation}   
\begin{equation}
\phi(t) = \int_{-\infty}^{t}{v(\tau)d\tau}
\end{equation}
\begin{equation}
q(t) = \int_{-\infty}^{t}{i(\tau)d\tau}
\end{equation}
where $\phi(t)$ is flux linkage, $q(t)$ is the amount of electric charge flowed through the memristor, $v(t)$ is voltage and $i(t)$ is current. Hence, the memristance of the device could be defined as shown in equations 7-8.
\begin{equation}
M(q) = \frac{d\phi}{dq}
\end{equation}
\begin{equation}
M(q(t)) = \frac{\frac{d\phi}{dt}}{\frac{dq}{dt}} = R_M(q)
\end{equation}
Finally, it could be concluded for memristance (equation 9):
\begin{equation}
v(t) = M(q(t))\ i(t)
\end{equation}
Hence, I-V plot of memristor could be sketched (refer to figure 1).
\begin{figure}[h]
\includegraphics[scale=0.45]{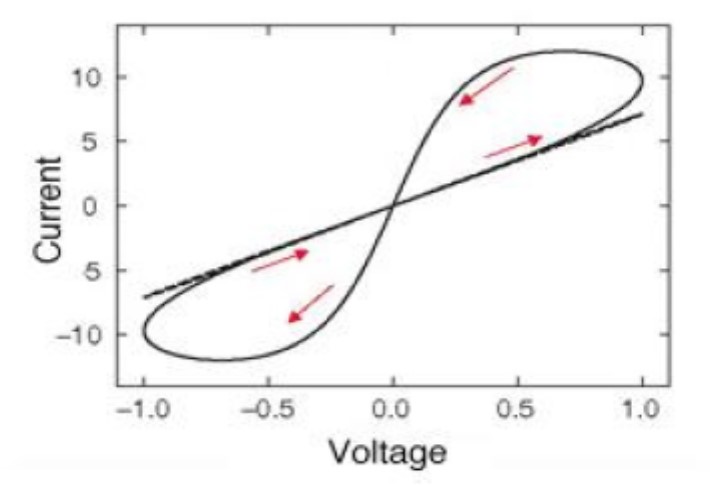}
\caption{Current-Voltage characteristics of a memristor (from {\cite{5}})}
\end{figure}\\
In addition, scientists recommend to implement memristors to the design of filters, since they allow to create electronically adjustable filters {\cite{9}}. 
\section{Simulations and Mathematical Analysis}
\subsection{First Approach}
The first step was to find appropriate electrical circuit for a filter with transfer function of Gaussian. Initially, lumped-element ladder-network approximation of the complete Gaussian transmission line (refer to figure 2) was considered. 
\begin{figure}[h]
\includegraphics[scale=0.5]{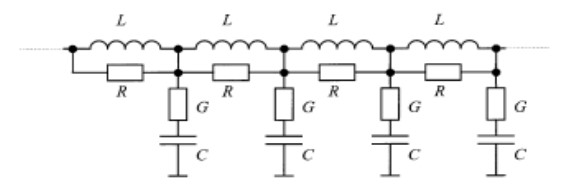}
\caption{Lumped-element ladder-network approximation of the complete Gaussian transmission line (from {\cite{2}})}
\end{figure}
This circuit (figure 2) was constructed on Spice (figure 3) and simulated. The values for passive elements were selected approximately as follows: $R_1=R_5=0.5\ k\Omega$,\ $R_2=R_3=R_4=1\ k\Omega$, \ $R_6=R_7=R_8=R_9=2\ k\Omega$, \ $C_1=C_5=0.6\ \mu F$, $C_2=C_3=C_4=1.2\ \mu F$, $L_1=L_2=L_3=L_4=0.1\ H$. The result of the simulation is shown in figure 4.
\begin{figure}[h]
\includegraphics[scale=0.28]{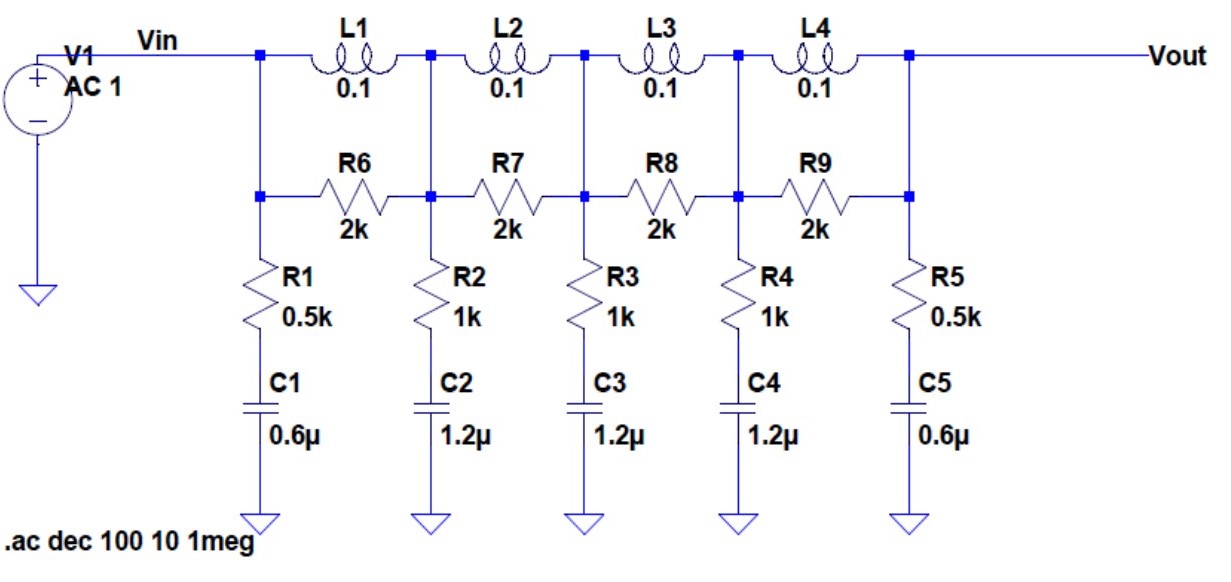}
\caption{Constructed on Spice circuit of ladder-network}

\includegraphics[scale=0.28]{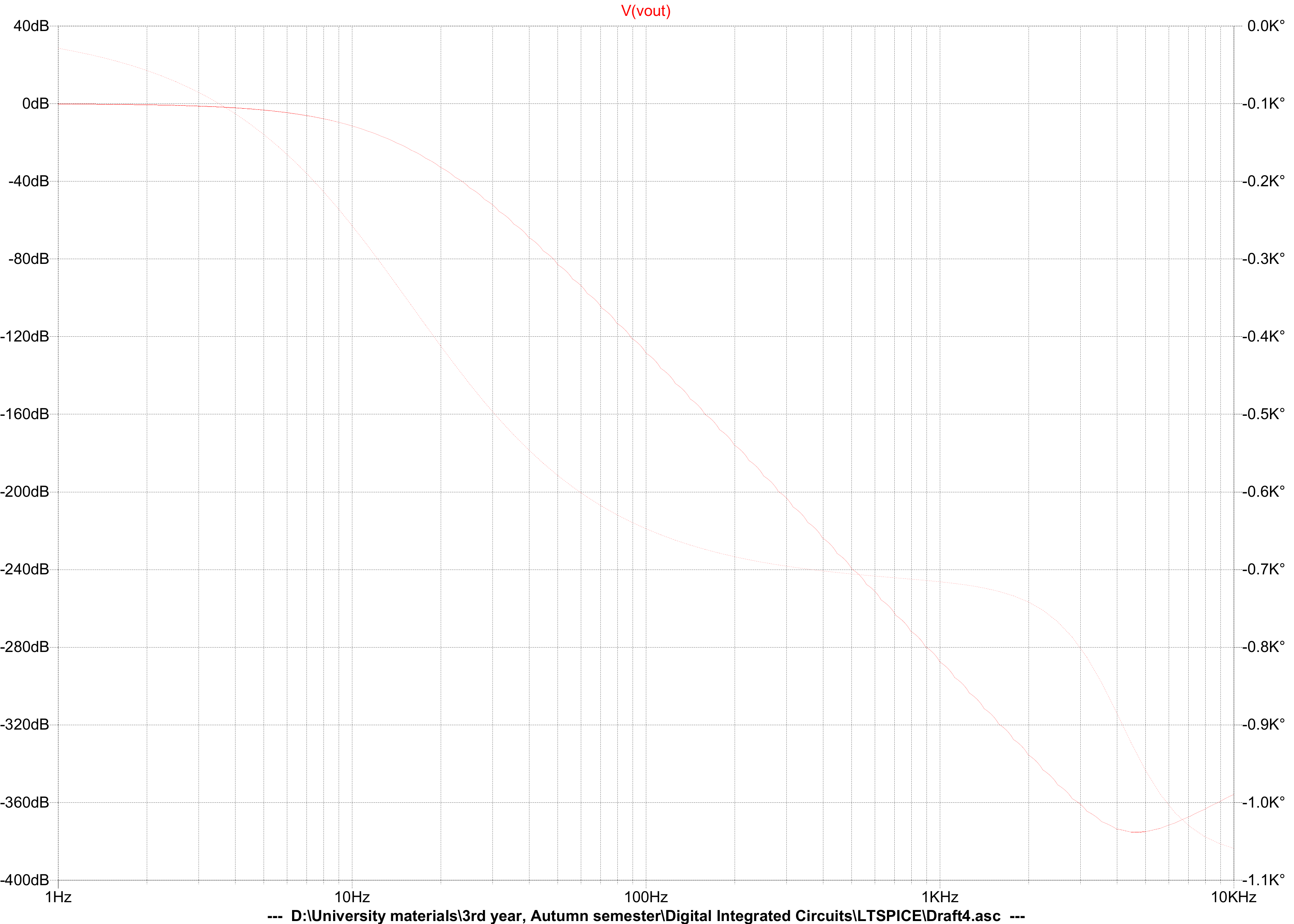}
\caption{Spice simulation of the ladder-network}
\end{figure}\\
The sketched solid curve in figure 4 is a magnitude response of the designed filter. The behaviour of this curve is close to the behaviour of Gaussian. However, in this electrical circuit inductors were used {\cite{8}}. It is recommended {\cite{7}} to avoid the use of inductors for the design of electrical filters due to size, width, lack of isolation from environment, energy losses and magnetic field issues. Therefore, another design for the filer was considered.

As it was mentioned before, the circuit shown in figure 3 was considered as the approximation of Gaussian-like filter. Here, input signal is $V_{in}$ and output signal - $V_{out}$. The transfer function of the filter could be considered, since it provides an algebraic representation of LTI filter in the frequency domain. The formula for transfer function is given as $H(z)=Y(z)/X(z)$, where $Y(z)$ is output, $X(z)$ is input and $H(z)$ stands for transfer function, which is the $z-transform$ of the impulse response $h(n)$. On spice this result could be represented as the ratio of the output voltage to the input voltage, i.e., $V_{out}/V_{in}$ and the result is shown on figure 4. As can be seen above, the transfer function of the system considered have the properties akin to the properties of an ideal Gaussian filter. As it was expected, this filter operates as low-pass filter, since it attenuates high frequencies.
	The next step, it to find approximate equation for the obtained transfer function. This could be done via Excel, by inserting all data points and plotting. For simple approximation 20 data points in the of the transfer function is going to be considered. The collected data was reported to table 1 below.
\begin{table}[h]
\caption{Reported Values}
\centering
\begin{tabular}{|l|p{2.5cm}|p{2.5cm}|}
\hline
\# & Frequency ($Hz$) & Magnitude ($dB$)\\
\hline
1&10000&0.000\\
\hline
2&20000&0.136\\
\hline
3&40000&0.516\\
\hline
4&100000&2.496\\
\hline
5&174480&3.969\\
\hline
6&236560&2.585\\
\hline
7&297230&0.028\\
\hline
8&397630&-4.200\\
\hline
9&615770&-11.130\\
\hline
10&1000&-20.060\\
\hline
11&2000&-32.010\\
\hline
12&3000&-38.080\\
\hline
13&4000&-41.410\\
\hline
14&5000&-43.640\\
\hline
15&6000&-44.870\\
\hline
16&7000&-45.770\\
\hline
17&8000&-46.340\\
\hline
18&9000&-47.020\\
\hline
19&10000&-48.450\\
\hline
20&15000&-48.460\\
\hline
\end{tabular}
\end{table}
According to these data the mathematical equation (equation 10) has been obtained.
\begin{equation}
y = -9.087\ ln(x) + 38.758
\end{equation}
In this equation $x$ stands for the frequency and $y$ stands for the magnitude in decibels. However, this approximation is not quite accurate. As the number of points considered increase, the accuracy of the obtained equation also increases.

\subsection{Second Approach}
The electrical circuit designed before was changed to the design, known as an 8-order Sallen-Key filter, which is considered as the approximation to Gaussian filter. Then, all resistors were replaced with memristors and the resultant circuit is shown in figure 5.

\begin{figure}[h]
\includegraphics[scale=0.29]{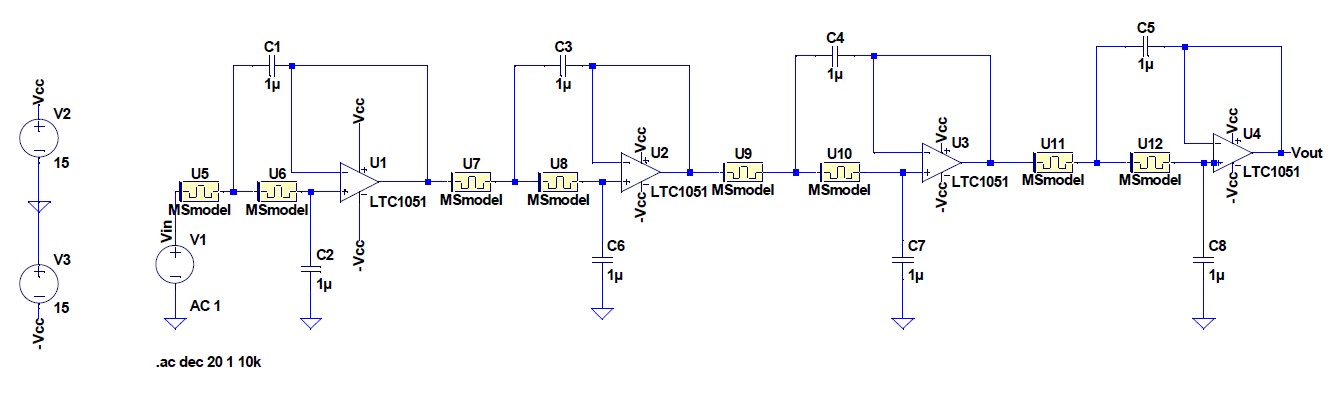}
\caption{8-order memristor-based approximation to Gaussian filter}
\end{figure}\

\section{Results}
The magnitude-frequency response of this designed filter is shown in figure 6. 
\begin{figure}[h]
\includegraphics[scale=0.30]{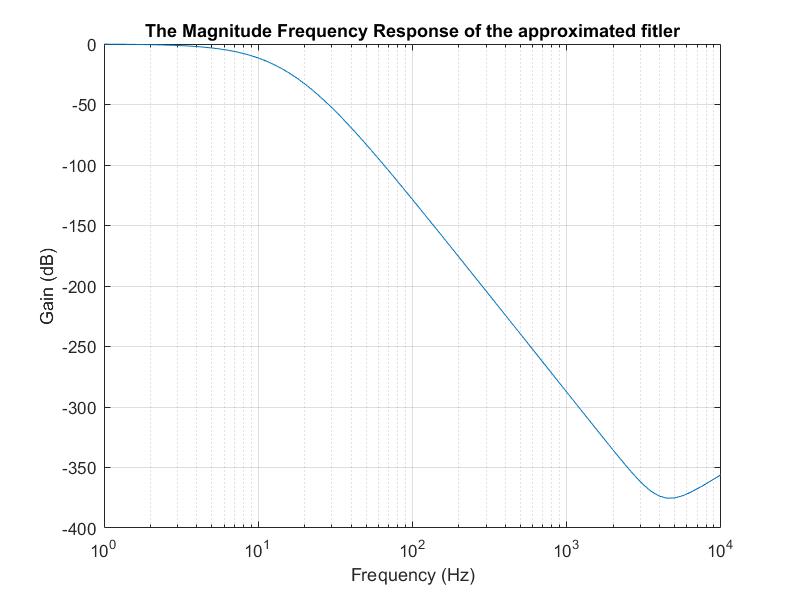}
\caption{Magnitude-frequency response of an 8-order Gaussian approximation filter}
\end{figure}\

The cutoff frequency is $f_c = 4.78 Hz$ and phase shift is $-135\ deg$. This  result satisfies to expected theoretical approaches. Now the transfer function of the filter could be determined according to the equation 3. Hence, the transfer function of this filter is:
\begin{equation}
H(s) = \frac{0.2 s^3 + 1.2 s^2 - 5 s + 40}{2.9 s^4 + 12 s^3 + 27.4 s^2 + 34 s + 19}
\end{equation}. 
\section{Conclusion}
In this paper Gaussian analog filter based on memristor was approximated. Firstly, Gaussian-like filter was designed from lumped-element ladder-network. However, due to the presence of inductors, this design was changed to Sallen-Key filter design. Finally, an 8-order memristor-based Gaussian filter was designed and its parameters were determined. The advantages of this design are that this circuits does not contain inductors and could be constructed simply with RLC elements. Furthermore, the use of memristors significantly increases the sensitivity of this filter. Nevertheless, not all aspects of memristors were considered. For instance, since the resistance of a memristor changes according to input voltage, this filter should be simulated at different inputs. Hence, it is recommended to focus further studies on memristor's resistance to input voltage relation and its effect to cutoff frequency, because ability to adjust the cutoff frequency and gain are crucially important.

\end{document}